\magnification 1200

\font\titlefont=cmss17 at 29.88 true pt
\font\authorfont=cmssi17 at 20.74 true pt
\font\bigaddressfont=cmss12 at 14.4 true pt 
 at 14.4 true pt
\font\bigmathi=cmmi12 at 27.4 true pt
\font\bigmathsy=cmsy12 at 24.4 true pt
\font\bigmathex=cmex12 at 24.4 true pt
\font\bigtenrm=cmr10 at 20.4 true pt
\font\bigtwelverm=cmr12 at 24.4 true pt
\def \mysubmit {}
\def \mypresent {}
\def\bigmath{\textfont1=\bigmathi \textfont2=\bigmathsy \textfont3=\bigmathex
          \textfont0=\bigtwelverm  
\scriptfont0=\bigtenrm}
\def \docnum #1 { \def \mydocnum {#1}} 
\def \date #1 { \def \mydate {#1}} 
\def \title #1 {\def \mytitle {#1}}
\def \author #1 {\def \myauthor {#1}}
\def \abstract #1 {\def \myabstract {#1}}

\def \tobesubmittedto #1 { \def \mysubmit {\leftskip=0pt plus 1fill \rightskip=0pt plus 1fill 
\hbox{\vbox{\noindent \hfill \it To be 
submitted To #1 \hfill \ }}}}
\def \submittedto #1 { \def \mysubmit {\leftskip=0pt plus 1fill \rightskip=0pt plus 1fill 
\hbox{\vbox{\noindent \hfill \it Submitted 
To #1 \hfill \ }}}}
\def \presentedat #1 { \def \mypresent {\leftskip=0pt plus 1fill \rightskip=0pt plus 1fill 
\hbox{\vbox{\noindent \it Presented 
at #1  }}}}

\def\maketitle{
\let\footnotesize\small
\let\footnoterule\relax

\ifx\mydate\undefined \def \mydate {
\ifcase\month\or
January\or February\or March\or April\or May\or June\or
July\or August\or September\or October\or November\or December\fi
\space\number\day, \number\year} \fi

\ifx\thispagestyle\undefined \nopagenumbers \fi
\ifx\nopagenumbers\undefined {\thispagestyle{empty}} 
      \setcounter{page}{0}%
      \fi
\null
\vskip 20 pt
\rightline{\logo}
\vskip 2 pt
\rightline{\mydocnum}
\rightline{\mydate}
\vskip 20 pt
{\def\\{\break} 
\leftskip=0pt plus 1fill
\rightskip = 0 pt plus 1fill
\parindent 0 pt
\baselineskip 30 pt
\bigmath
\titlefont \hfil \vbox { \mytitle}\hfil }
\vskip 25 pt
{\def \and {\qquad} 
\leftskip=0pt plus 1fill
\rightskip = 0 pt plus 1fill
\parindent 0 pt 
\authorfont
\lineskip 12 pt
\myauthor
\parfillskip=0pt\par
}%
\vskip 20 pt

{\bigaddressfont
\centerline{ Department of Physics}
\centerline{ Manchester University}
\centerline{ England}
}

\ifx\myabstract\undefined {}
\else
\null\vfil\vskip 10 pt
\centerline{ \bf Abstract}
\vskip 10 pt
\myabstract
\fi

\ifx\footline\undefined   
\begin{figure}[b]
\mypresent
\mysubmit
\end{figure}
\mythanks
\setcounter{footnote}{0} 
\vfil
\null
\else                       
\footline={{\baselineskip=10 pt \vbox{\hbox to \hsize {\mypresent} \vskip 5 pt \hbox to \hsize{\mysubmit}}}}
\vfil
\eject
\pageno=1 
\footline={\hss\tenrm\folio\hss}
\fi
				      
\let\thanks\relax
\gdef\mysubmit{}\gdef\present{}
\gdef\mythanks{}\gdef\myauthor{}\gdef\@title{}\let\maketitle\relax}

\def \phonenumber{4178}
\def\today{\number\day/\number\month/\number\year\space\number\hour%
:\number\minute\space\jobname}

\def \logo {\vbox to 23.5 mm {
\hbox {}
\hbox to 47 mm 
{
\includegraphics{/home/roger/tex/ulogo.ps} 
\hfil} \vfil }}

\def\header #1\par{ \centerline{\it #1}\par}
\def \letterhead {
\voffset -10 mm
{\advance \hsize by 2 cm
\font\address=cmss12
\vskip -9.5 pt
{\address
\vbox {
\vskip 2 mm
\hbox to 6 cm {\hskip -1 cm Department of Physics and Astronomy\hfill }
\hbox to 6 cm {\hskip -1 cm The University of Manchester \hfill }
\hbox to 6 cm {\hskip -1 cm Manchester \hfill }
\hbox to 6 cm {\hskip -1 cm M13 9PL \hfill }
\hbox to 8 cm {\hskip -1 cm Tel 0161-275-\phonenumber\hskip 1.0 cm 
Fax 0161-273-5867\hskip 1.0 cm}
\hfill
\vbox{
\hbox to 6 cm{\includegraphics{/home/roger/tex/ulogo.ps}} 
\vskip 0.5 mm
}
}
}}
}

\def \letter #1 {
\topskip 0 pt
\vsize 599 pt
\nopagenumbers
\letterhead
\vskip 1 mm

\vbox to 3.4 cm {\vfill #1 \vfill}
\vskip 5 mm

\hbox to 2 cm {\hskip -1 cm \leaders\hrule height .5 pt \hfill \hskip 1 cm}

\rightline {\number\day \
\ifcase\month\or January\or February\or March\or April\or
May\or June\or July\or August\or September\or October\or
November\or December\fi \ \number\year}

\vskip 5 mm
}

\def \AddressPPARC #1 { \leftline{#1 }
\leftline{PPARC,}\leftline{Polaris House,}
\leftline{North Star Avenue,}\leftline{SWINDON,}\leftline{SN2 1SZ}}

\def \AddressRAL #1 { \leftline{#1 }
\leftline{HEP Division,}\leftline{Rutherford Appleton Laboratory,}
\leftline{Chilton,}\leftline{Didcot,}\leftline{Oxon}}

\def \AddressRegistrar  #1 {
\leftline{#1 }
\leftline{The Registrar's Department}
\leftline{Main Building}
\leftline{University of Manchester}
\leftline{Oxford Road}
\leftline{Manchester M13 9PL}}

\def \AddressFaculty #1 {
\leftline{#1 }
\leftline{The Faculty of Science}
\leftline{Roscoe Building}
\leftline{University of Manchester}
\leftline{Oxford Road}
\leftline{Manchester M13 9PL}}

\def \AddressPhysics #1 {
\leftline{#1 }
\leftline{Department of Physics}
\leftline{University of Manchester}
\leftline{Oxford Road}
\leftline{Manchester M13 9PL}}

\def \AddressCERN #1/#2 {
\leftline {#1}
\leftline {#2 Division,}
\leftline {CERN,}
\leftline {CH1211 Gen\`eve 23,}
\leftline {Switzerland}}

\def \NIM #1 {{\it Nucl. Instr \& Meth. \/}{\bf A#1}\ }
\def \ZPC #1 {{\it Zeit. Phys. \/}{\bf C#1}\ }
\def \NPB #1 {{\it Nucl. Phys. \/}{\bf B#1}\ }
\def \PLB #1 {{\it Phys. Lett. \/}{\bf B#1}\ }
\def \PL #1 {{\it Phys. Lett. \/}{\bf #1}\ }
\def \PRD #1 {{\it Phys. Rev. \/}{\bf D#1}\ }
\def \PRL #1 {{\it Phys. Rev. Lett.\/}{\bf #1}\ }
\def \PR #1 {{\it Phys. Rev. \/}{\bf #1}\ }
\def \CPC #1 {{\it Comp. Phys. Comm. \/}{\bf #1}\ }

\newcount \eqnumber
\newcount \fignumber
\newcount \highref

\def \eq #1{\global\advance \eqnumber by 1
\let \rrr=\eqnumber
\xdef #1{\the\rrr}
\eqno (\the\eqnumber)
}
\def \fig #1{Figure \global\advance \fignumber by 1 \the\fignumber
\let \rrr=\fignumber
\xdef #1{Figure \the\sss}
}

\newcount\refcheck
\newcount\thisrf
\def\references #1 {
\thisrf=0
\ifnum\refcheck=0
\else
\leftline{\bf References}
\fi
\input #1
\refcheck=1
}

\def\refiopstyle #1:#2;#3\par{
\relax
\ifnum\refcheck=0
\edef \rrr{#2}
\let #1=\rrr
\else
#2 
\ 
#3
\par
\fi
\relax
}

\def\refseq#1{\ifnum#1>\the\highref\global\advance\highref 1  
\ifnum#1>\the\highref
\message{Reference out of Sequence: expecting \the\highref got #1}
\highref=#1\fi\fi}

\def\reference #1:#2\par{\advance \thisrf by 1
\relax
\ifnum\refcheck=0
\let \sss=\thisrf
\edef \rrr{\the\thisrf\noexpand\refseq{\the\thisrf}}
\let #1=\rrr
\else
\the\thisrf
:\ 
#2
\par
\fi
\relax
}

\font\small=cmr8

\def \tickbox {\lower 2 mm \hbox{
\vbox  {\vskip .5 mm\hrule \hbox to 4 mm
{\vrule \strut  \hfill  \vrule}\vfill\hrule} }}

\newcount\secnum
\newcount\examplenum
\newcount\subnum
\newcount\subsubnum
\newcount\chapnum
\font \splash=cmssi17
\font \titlefont=cmss17 scaled \magstep2

\font \address=cmss12
\font \cf=cmbxsl10
\examplenum=0
\secnum=0
\subnum=0

\def\chapter #1 {
\advance\chapnum by 1 \secnum=0 \subnum=0 \subsubnum=0
\vfill
\hbox{\bf \quad Chapter \number \chapnum : #1}}
\def\lecture #1 #2{
\secnum=0 \subnum=0
\hbox{\centerline{\splash SLUO Lecture #1: #2}}
\headline={\ifnum\pageno>1 SLUO Lecture #1 \dotfill #2 \fi}
\footline={\ifnum\pageno>1 \hss --\  \folio \ -- \hss  \else  \fi}
}

\def\subsection#1 \par{\par \advance\subnum by 1
\subsubnum=0
\goodbreak \vskip 0.3cm\leftline{\cf \number \secnum .\number 
\subnum \ #1} \par}

\def\subsubsection#1 \par{\par \advance\subsubnum by 1
\goodbreak \vskip 0.3cm\leftline{\sl \number \secnum .\number 
\subnum .\number \subsubnum \ #1} \par}

\def\section#1\par{\goodbreak \par \advance\secnum by 1 \subnum=0
\vskip 0.5cm\leftline{\bf \number \secnum . \ #1}\vskip 0.02cm \par
\message{ Section  \number \secnum    #1}
}

\long\def\example #1 {\par \advance\examplenum by 1
\vskip 12 pt \goodbreak \boxit {{\bf Example \number\examplenum :} #1 }
}

\long\def\boxit #1{\vbox {\kern-5pt\hrule\hbox{\vrule\kern3pt
           \vbox{\kern3pt #1 \kern3pt} \kern3pt \vrule} \hrule}}

\def\bull#1\par {\item {$\bullet$} #1 \par}


\title {A Note on $\Delta ln L=-{1 \over 2}$ Errors}
\docnum {MAN/HEP/04/02}
\date{1/3/2004}
\author {Roger Barlow}
\abstract {
The points at which the log likelihood falls
by ${1 \over 2}$ from its maximum value are often used to give the `errors' on a
result, i.e. the 68\% central confidence interval. The validity of
this is examined for two 
simple cases: a lifetime measurement and a Poisson measurement.
Results are compared with the exact Neyman construction and with
the simple Bartlett approximation.  It is shown that the accuracy
of the log
likelihood method is poor, and the Bartlett construction explains why
it is flawed.  
}

\maketitle

\section Introduction

In the limit where the number of measurements $N$ is large, the variance of the 
 maximum
likelihood estimator $\hat a$ of a parameter $a$ is given by
$$V(\hat a)=\left( - {d^2 ln L \over da^2}\right)^{-1} \eqno(1)$$
and the quoted error $\sigma_{\hat a}=\sqrt{V(\hat a)}$
can be read off the parabolic
likelihood curve from the points at which the likelihood $L(a)$
falls by ${1\over 2}$ from its peak value $L(\hat a)$:
$\Delta ln L=-{1 \over2}$.

For experiments with finite $N$ a similar procedure is in general use:
the values $a_\pm$ below and above $\hat a$ for which $\Delta ln L=ln L(a_\pm)-
ln L(\hat a)=-{1 \over 2}$ are found, and the 68\% central
confidence interval quoted as $[a_-,a_+]$ or $[\hat a - \sigma_-,
\hat a + \sigma_+]$.

This is given a somewhat non-rigorous justification [1,2,3]: 
even though the log likelihood curve for $a$ may not be a parabola,
the parameter $a$ could be converted to some $a'$ for which the
log likelihood curve is parabolic; 
symmetric errors $\sigma_{a'}$ could be read off in
the standard way, and the $a'$ interval converted back to the corresponding
interval for $a$.  The invariance of the maximum likelihood formalism then
ensures that this interval is just the 
$\Delta ln L=-{1 \over2}$ 
interval 
for $a$. 

 This practice is now being questioned [4,5,6] and
an examination of how well it actually works in practice is needed to
inform this discussion.
In this note we consider two typical cases where Maximum Likelihood
estimation is used: the determination of the lifetime of an unstable state
decaying according to the radioactive decay law, and the determination
of the number of events produced by a Poisson process.  In these we can
determine the interval produced by the $\Delta ln L=-{1 \over 2}$
recipe and contrast them with the exact Neyman interval.  This is
found [2,7] from the values satisfying:

$$\eqalign {\int_0^{\hat a} P(\hat a';a_+) d\hat a' = 0.16\cr
\int_{\hat a}^\infty P(\hat a';a_-) d \hat a' = 0.16}\eqno(2)
$$
where $P(\hat a;a)$ is the probability density for a true value 
$a$ giving an estimate $\hat a$.  These equations define the 
confidence belt such that the probability of a
measurement lying within the region is, by construction, 68\%.

An alternative approximation technique is that of Bartlett [1,7,8].
For any $N$ the quantity ${d ln L \over da}$ is distributed with mean zero and
variance $-\left< {d^2 ln L \over da^2} \right>$.
For large $N$ the Central Limit Theorem
prescribes that ${d ln L \over da} = \sum_1^N {d ln P(x_i;a) \over da}
$, the sum of $N$ random quantities, 
is Gaussian.
If this quantity can be expressed in terms of $\hat a - \left< \hat a \right>$
this can be used to give confidence regions for $\hat a$.
Further refinements can be used to correct for the non-Gaussian finite
$N$ behaviour, but these lie beyond the scope of this work.

This note uses the 68\% central confidence region for illustration, but the 
techniques can be applied to central or one-sided regions with any 
probability content.

Bayesian statistics can also be used to give confidence intervals.
This is an entirely different techique, and is not considered here.
This study compares the exact Neyman confidence intervals with two
methods which claim to approximate to them.

\section Lifetime Measurements

The probability for a state with mean lifetime $\tau$ to decay after
an observed time $t$ is given by
$$P(t;\tau)={1 \over \tau} e^{-t/\tau}.\eqno(3)$$
The log likelihood for $N$ measurements $t_1 \dots t_N$ is
\def \tbar {{\overline t}}
$$ln L = - N {\tbar \over \tau} - N ln \tau \eqno(4)$$
where $\tbar = {1 \over N} \sum t_i$. Differentiation to find the maximum
immediately gives $\hat \tau=\tbar$ and $ln L(\hat \tau)=-N(1+ln \tbar)$.
The problem scales with $\tau /\tbar$, and without loss of generality
we can take $\tbar = 1.$  We consider the 68\% confidence region
for various values of $N$.

The probability of obtaining a particular value of $\tbar$ contains a
term $e^{-N \tbar/\tau}$ from equation 3, and a factor $\tbar^{N-1}$
from the convolution.  Normalisation gives (see [5], Equation 4)
$$P(\tbar;\tau)={N^N \tbar^{N-1} \over \tau^N (N-1)!} e^{-N \tbar / \tau}. \eqno(5)$$
For the exact Neyman region we require the integral of this quantity
from zero to the measured value, which is to be 16\% for the upper limit
$\tau_+=\tbar+\sigma_+$ and 84\% for the lower limit $\tau_-=\tbar-\sigma_-$. This is given by
$$\int_0^\tbar P(\tbar';\tau) \, d \tbar' 
= 1-e^{-N \tbar/\tau} \sum_{r=0}^{N-1} {\tbar^r N^r \over r! \tau^r}. \eqno(6)
$$

The region thus obtained, expressed as differences from the measured $\tbar$
of 1, is shown in the columns 2 and 3 of Table 1, for values between $N=1$ to $N=25$.

\vskip\baselineskip
\vbox{
\centerline{\vbox{\halign{
# & # & # &\qquad # & # &\qquad # & #   \cr
N& Exact & & \hfil$\Delta \ln L$&$=-{1 \over 2}$\hfil& Bartlett & \cr
&
\hfil $\sigma_-$ \hfil &
\hfil $\sigma_+ $\hfil &
\hfil $\sigma_-$ \hfil &
\hfil $\sigma_+ $\hfil &
\hfil $\sigma_-$ \hfil &
\hfil $\sigma_+ $\hfil 
\cr
   1 &  0.457&  4.787 & 0.576  &2.314 & 0.500 &  $\infty$ \cr
   2 &  0.394&  1.824 & 0.469  &1.228 & 0.414 &2.414\cr
   3 &  0.353&  1.194 & 0.410 & 0.894 & 0.366 &1.366\cr
   4 &  0.324& 0.918 & 0.370 &0.725 & 0.333&1.000\cr
   5 &  0.302& 0.760 & 0.340 &0.621 & 0.309&0.809\cr
   6 &  0.284& 0.657 & 0.318 &0.550  & 0.290 &0.690\cr
   7 &  0.270& 0.584 & 0.299 &0.497 & 0.274&0.608\cr
   8 &  0.257& 0.529 & 0.284 &0.456 & 0.261&0.547\cr
   9 &  0.247& 0.486 & 0.271 &0.423 & 0.250&0.500\cr
  10 &  0.237& 0.451 & 0.260 &0.396 & 0.240&0.463\cr
  15 &  0.203& 0.343 & 0.219 &0.310  & 0.205&0.348\cr
  20 &  0.182& 0.285 & 0.194 &0.261 & 0.183&0.288\cr
  25 &  0.166& 0.248 & 0.176 &0.230  & 0.167&0.250\cr
}}}
\vskip\baselineskip
Table 1: 68\% Confidence regions obtained by the 3 methods for a lifetime
measurement 
\vskip\baselineskip
}

The $\Delta ln L = - {1 \over 2}$ points can be found numerically from Equation 4.  These are shown in columns 4 and 5 of Table 1.

For the Bartlett approximation, the differential of Equation 4 gives 
${N \over \tau^2} (\tbar - \tau)$, and the expectation value of the
second differential gives the variance of this as ${N \over \tau^2}$.
Thus for a given $\tau$ the probability distribution for $\tbar$
has mean $\tau$ and standard deviation $\tau / \sqrt N$.
This is exact. We then  -- this is the approximation -- 
take this as being Gaussian and use it in the Neyman prescription, accordingly 
requiring that $\tbar$ lie one standard deviation above $\tau_-=\tbar-\sigma_-$ and one
standard deviation below $\tau_+=\tbar+\sigma_+$

$$ \tbar = \tau_- + {\tau_- \over \sqrt N} \qquad 
\tbar = \tau_+ - {\tau_+ \over \sqrt N}\eqno(7)$$
i.e. $\sigma_- = {\tbar \over  \sqrt N+1 }$ and $\sigma_+ ={\tbar \over  \sqrt N-1}$.
These are shown in the final two columns of Table 1.
The results are also presented graphically in Figure 1.

\vbox{
\vskip 9 cm
\includegraphics{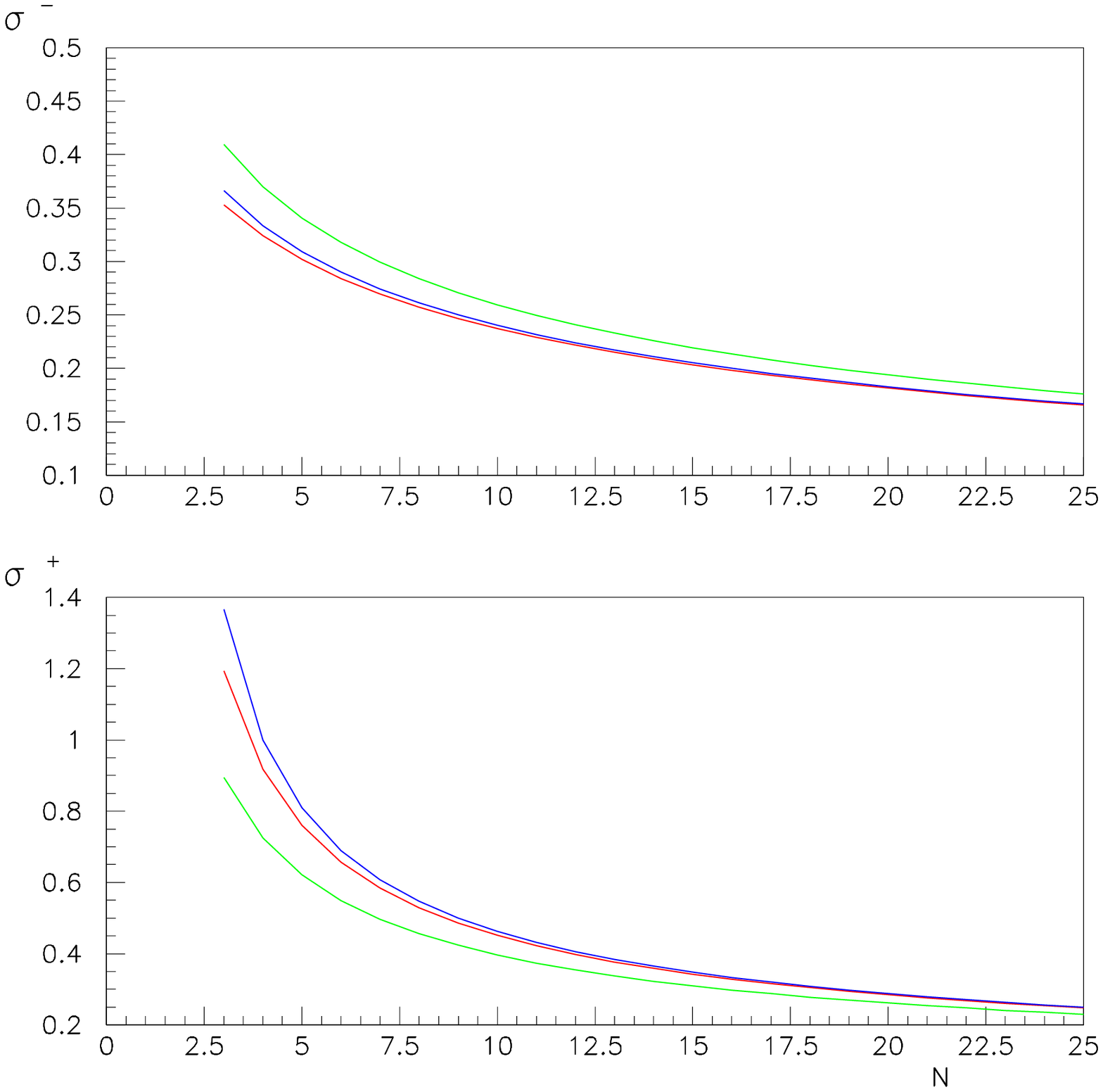}
\centerline{\vbox{Figure 1: 
Upper and lower limits on the 68\% central confidence interval for a lifetime
measurement showing the exact construction (red), the Bartlett approximation (blue) and the $\Delta \ln L$ approximation (green)
}}
}
\vskip\baselineskip

Two points emerge, from both Table 1  and Plot 1. One is that the
Bartlett approximation does surprisingly well (except at very  small $N$,
 of  order 1).  The second is that the Log likelihood approximation does surprisingly badly.  
For $N\sim 10$ the differences are of order 10\%. 
The convergence towards agreement is clearly slow.

\vfill\eject

\section Poisson Measurements

If $N$ events are seen from a Poisson process, Equation 2 gives
the upper and lower limits of the 68\% central region as
$$\sum_0^n e^{-\lambda_+}{\lambda_+^N\over N!}=0.16\qquad 
\sum_0^{n-1} e^{-\lambda_-}{\lambda_-^N\over N!}=0.84.\eqno(8)$$
These are shown in columns 2 and 3 of Table 2 for a range of values of $N$. The
$\Delta ln L=-{1 \over 2}$
errors are read off 
$N-\lambda + N ln (\lambda/N)$.
These are shown in columns 4 and 5 of Table 2.

\vskip\baselineskip
\vbox{
\centerline{\vbox{\halign{
# & # & #\qquad & # &  # \qquad & # & #   \cr
N& Exact & &\hfil $\Delta \ln L$&$=-{1 \over 2}$\hfil& Bartlett & \cr
&
\hfil $\sigma_-$ \hfil &
\hfil $\sigma_+ $\hfil &
\hfil $\sigma_-$ \hfil &
\hfil $\sigma_+ $\hfil &
\hfil $\sigma_-$ \hfil &
\hfil $\sigma_+ $\hfil 
\cr
   1 &   0.827 &  2.299 &  0.698 &  1.358 &   1.118 & 2.118\cr
   2  &   1.292 &  2.637 &   1.102 &  1.765 &   1.500 & 2.500\cr
   3  &   1.633 &  2.918 &   1.416 &  2.080 &   1.803 & 2.803\cr
   4  &   1.914 &  3.162 &   1.682 &  2.346 &   2.062 & 3.062\cr
   5  &   2.159 &  3.382 &   1.916 &  2.581 &   2.291 & 3.291\cr
   6  &   2.380 &  3.583 &   2.128 &  2.794 &   2.500 & 3.500\cr
   7  &   2.581 &  3.770 &   2.323 &  2.989 &   2.693 & 3.693\cr
   8  &   2.768 &  3.944 &   2.505 &  3.171 &   2.872 & 3.872\cr
   9  &   2.943 &  4.110 &   2.676 &  3.342 &   3.041 & 4.041\cr
  10  &   3.108 &  4.266 &   2.838 &  3.504 &   3.202 & 4.202\cr
  15  &   3.829 &  4.958 &   3.547 &  4.213 &   3.905 & 4.905\cr
  20  &   4.434 &  5.546 &   4.145 &  4.811 &   4.500  &5.500\cr
  25  &   4.966 &  6.066 &   4.672 &  5.339 &   5.025 & 6.025\cr
}}}
\vskip\baselineskip
Table 2: 68\% Confidence regions obtained by the 3 methods for a Poisson
measurement 
\vskip\baselineskip
}

The Bartlett method gives the familiar fact that the variance
of $n-\lambda$ is just $\lambda$.  This suggests that
$$n-\lambda_- = \sqrt \lambda_- \qquad \lambda_+ - n=\sqrt \lambda_+.$$
However $P(n;\lambda)$ is defined for
integer $n$ only. To make this set of discrete spikes look 
like a Gaussian requires us to replace it by a histogram where
the value is defined as $\exp^{-\lambda} \lambda^n/n!$
for values of the continuous abscissa variable between
 $n-{1 \over 2}$
 and $n+{1 \over 2}$.  This  requires us to add ${1 \over 2}$ to
each of the ranges, giving
$$\sigma_-=\sqrt{n+{1 \over 4}}\qquad \sigma_+=\sqrt{n+{1 \over 4}}+1\eqno(9)$$
These are shown in columns 6 and 7  of Table 2.
The data are shown graphically in Figure 2.

\vbox{
\vskip 9 cm
\includegraphics{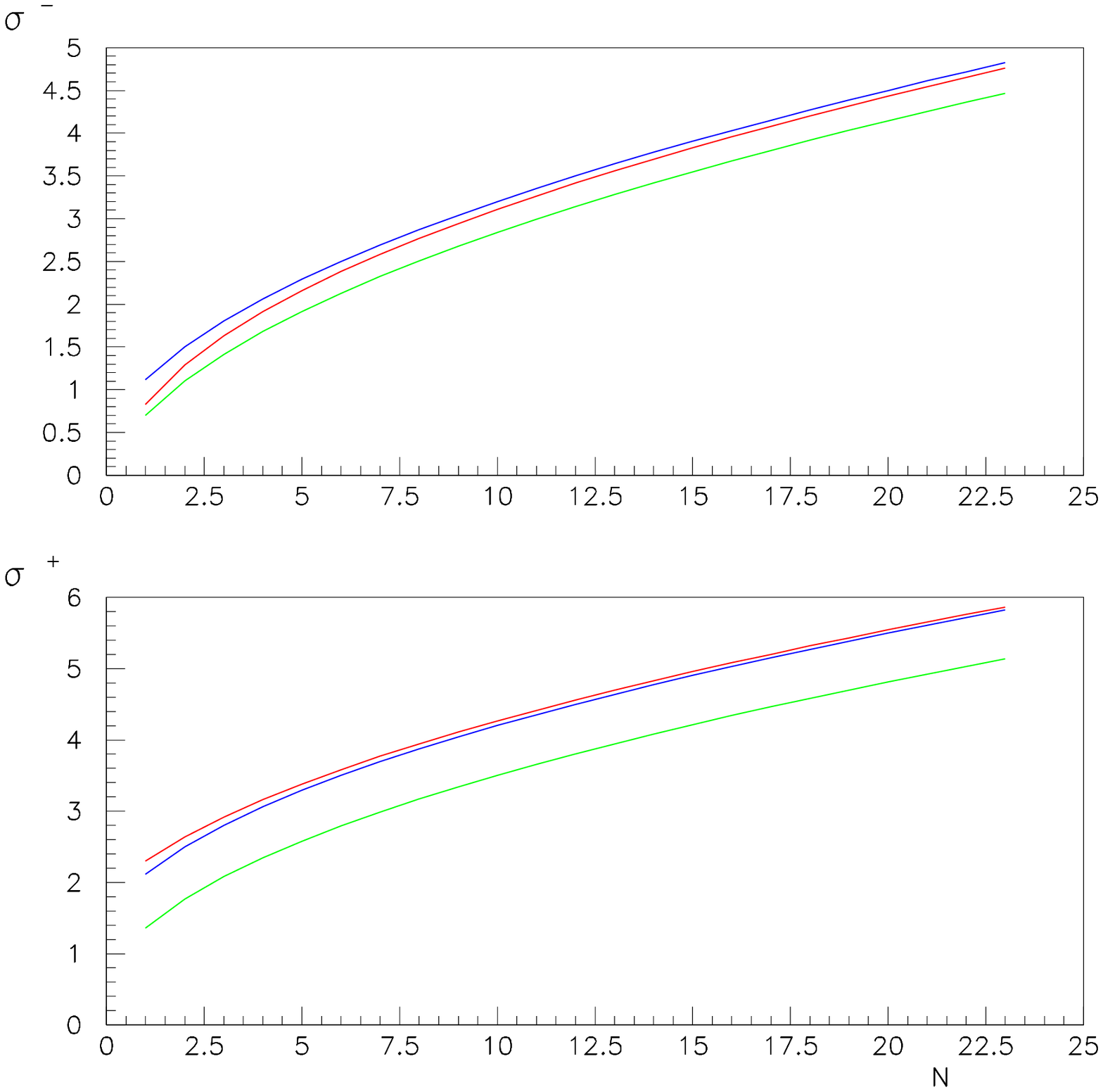}
\centerline{\vbox{Figure 2: 
Upper and lower limits on the 68\% central confidence interval for 
a Poisson measurement, showing the exact construction (red), the Bartlett approximation (blue) and the $\Delta \ln L$ approximation (green)
}}
}
\vskip\baselineskip

Again, the Bartlett approximation does surprisingly well, and the 
$ln L$ approximation surprisingly badly. 
Furthermore, in this case it underestimates both errors, which will
inevitably lead to a smaller than desired coverage. (This could be remedied by
adding 0.5 to each limit, to account for the discrete binning, though this
is still worse than the Bartlett approximation, as can be seen from Table 2.)

\section Summary

The poor behaviour of the log likelihood error approximation can be 
understood
within the Bartlett approximation.  The distribution for ${d \ln L \over d a}$
is re-expressed in terms of a distribution for $a - \hat a$ which is assumed to be Gaussian
$$p(\hat a ;a)={1 \over \sqrt {2 \pi} \sigma(a)} e^{-(a-\hat a)^2 / 2 \sigma(a)^2}\eqno(10)$$
where the notation $\sigma(a)$ makes the point that the variance of this
Gaussian depends on $a$.

The 68\% limits are given by finding the  $a$ for which 
$\hat a - a =\pm \sigma(a)$. These do indeed 
correspond to a fall of ${1 \over 2}$
in the log likelihood from the exponential.  However the total log likelihood
also changes with $a$ due to the $ -\ln \sigma(a)$ from the 
denominator.  The simple 
$\Delta \ln L=-{1 \over 2}$
 method considers
all factors together, and thus wrongly includes this term. 

The inaccurary of the logarithmic method is appreciable. For
reasonable values of $N$ it is generally wrong in the second 
significant figure, and often pretty grossly wrong. That this 
occurs for both cases examined suggests that this is true in
general. And yet values obtained by this method are frequently
quoted to considerable precision by experiments.

In the complicated likelihood 
functions used in real experimental results, a simple Bartlett
approach may not be possible.  However the logarithmic approximation 
clearly does not provide the accuracy with which experiments wish to
report their results.
An alternative, available today but not in the 1950's when these
techniques were developed, is to 
use the known Likelihood function to
perform the Neyman construction using Monte Carlo integration (the so-called `toy Monte Carlo'). 
This should be strongly recommended.

\vskip \baselineskip
\leftline{\bf Acknowledgements}

The author gratefully acknowledges the support of the Fulbright Foundation.

\vskip \baselineskip

\parindent 0 pt

\leftline{\bf References}

[1] A.G. Frodeson {\it et al.}: {\it Statistics for Nuclear and Particle Physicists}
Universitetsforlaget Bergen-Oslo-Tr\o mso, 1979.

[2] R.J. Barlow: {\it Statistics: A Guide to the Use of Statistical Methods in the Physical Sciences,} John Wiley \& Sons. 1989.

[3] W.T. Eadie {\it et al.}: {\it Statistical Methods in Experimental
Physics}, North Holland, 1971.

[4] R.J. Barlow:{\it Introduction to Statistical Issues in Particle Physics,}
\hfill\break
\qquad arXiv physics/0311105
To appear in proceedings of PHYSTAT2003, SLAC, 2004.

[5] A. Bukin:{\it A Comparison of Methods for Confidence Intervals,}
\hfill\break
\qquad arXiv physics/0309077
To appear in proceedings of PHYSTAT2003, SLAC, 2004.

[6] G. Zech, private communication.

[7] M.G. Kendall and A. Stuart: {\it The Advanced Theory of Statistics}, Charles Griffin \& Co.,
Vol II, 
4th Edition (1979)

[8] M.S. Bartlett: {\it On the Statistical Estimation of Mean Lifetimes,}
Phil. Mag. {\bf 44} 244 (1953),\hfill\break
--- \qquad  {\it Estimation of Mean Lifetimes from Multiple Plate Cloud Chamber Tracks,}
Phil. Mag. {\bf 44} 1407 (1953)

\bye